\documentclass[useAMS,referee,usenatbib]{biom}
\usepackage{mathtools}
\usepackage{bbm}
\usepackage{xcolor}

\def\bSig\mathbf{\Sigma}

\title[Quantile treatment effects with missing data]{Estimating weighted quantile treatment effects with missing outcome data by double sampling}

\author{Shuo Sun$^{1,*}$\email{shuosun@hsph.harvard.edu}, 
Sebastien Haneuse$^{1}$,
Alexander W. Levis$^{2}$, 
Catherine Lee$^{3}$,
David E Arterburn$^{4}$,\\
\textbf{Heidi Fischer$^{5}$},
\textbf{Susan Shortreed$^{4,6}$},
\textbf{and Rajarshi Mukherjee$^{1}$}\\
$^{1}$Department of Biostatistics, Harvard T.H. Chan School of Public Health, Boston, MA, USA \\
$^{2}$Department of Statistics \& Data Science, Carnegie Mellon University, Pittsburgh, PA, USA\\
$^{3}$Division of Research, Kaiser Permanente Northern California, Oakland, CA, USA\\
$^{4}$Health Research Institute, Kaiser Permanente Washington, Seattle, WA, USA\\
$^{5}$Department of Research and Evaluation, Kaiser Permanente Southern California, Pasadena, CA, USA\\
$^{6}$Department of Biostatistics, University of Washington, Seattle, WA}

\begin{document}


\date{{\it Received October} 2007. {\it Revised February} 2008.  {\it
Accepted March} 2008.}



\pagerange{\pageref{firstpage}--\pageref{lastpage}} 
\volume{64}
\pubyear{2008}
\artmonth{December}


\doi{10.1111/j.1541-0420.2005.00454.x}


\label{firstpage}


\begin{abstract}
Causal weighted quantile treatment effects (WQTEs) complement standard mean-focused causal contrasts when interest lies at the tails of the counterfactual distribution. However, existing methods for estimating and inferring causal WQTEs assume complete data on all relevant factors, which is often not the case in practice, particularly when the data are not collected for research purposes, such as electronic health records (EHRs) and disease registries.
Furthermore, these data may be particularly susceptible to the outcome data being missing-not-at-random (MNAR). This paper proposes to use double-sampling, through which the otherwise missing data are ascertained on a sub-sample of study units, as a strategy to mitigate bias due to MNAR data in estimating causal WQTEs. With the additional data, we present identifying conditions that do not require missingness assumptions in the original data. We then propose a novel inverse-probability weighted estimator and derive its asymptotic properties, both pointwise at specific quantiles and uniformly across quantiles over some compact subset of $(0,1)$, allowing the propensity score and double-sampling probabilities to be estimated. For practical inference, we develop a bootstrap method that can be used for both pointwise and uniform inference. A simulation study is conducted to examine the finite sample performance of the proposed estimators. We illustrate the proposed method using EHR data examining the relative effects of two bariatric surgery procedures on BMI loss three years post-surgery.
\end{abstract}

%



\maketitle


%

\section{Introduction}
Modern causal inference conceptualizes causation via a comparison of counterfactuals across interventions or treatment strategies \citep{hernan2020}. As causation for individuals cannot be directly observed, analysis typically focuses on contrasts of some summary statistic of the treatment-specific counterfactual distributions. The most common contrast, arguably, is the average treatment effect (ATE), although related contrasts have been considered including the ATE on the treated \citep[e.g.,][]{imbens2004nonparametric}, 
the survivor ATE \citep[e.g.,][]{ tchetgen2014identification}, 
and average effects of modified treatment policies \citep[e.g.,][]{munoz2012population, haneuse2013estimation}.
That each of these contrasts focuses on the mean of the counterfactual distribution, in one way or another, is natural since the mean is standard measure of centrality. In some settings, however, particularly when low-probability but high-impact outcomes are anticipated, understanding the causal effect of a given treatment beyond the `center', including at the tails of the counterfactual distributions, may be more relevant to decision-making \citep{sun2021causal, meyerson2023use}. In such settings, quantile treatment effects (QTEs), defined as the difference in quantiles of counterfactual distributions across treatment options, may be a useful compliment to standard contrasts based on the mean \citep[e.g.,][]{andrews2016quantile, autor2017effect, hemila2022quantile}. 

Towards estimation of causal QTEs, there is an emerging literature most of which focuses on inverse-probability weighting (IPW) using the propensity score \citep{firpo2007efficient, sun2021causal}. 
Throughout this literature, and in contrast to the literature on related but distinct quantile regression methods \citep{yuan2010bayesian, sherwood2013weighted, han2019general, yu2022quantile}, it has been implicitly assumed complete data are available the outcome and all relevant covariates. Missing or incomplete data, however, is a well-known and widespread practical challenge in almost all research settings involving humans. This is particularly the case when the data at-hand were not collected for research purposes as in the case with electronic health records (EHRs), claims-based databases and registries \citep{haneuse2017use}. In studies of the long-term effects of bariatric surgery, for example, data on relevant outcomes (e.g., weight/BMI, hypertension, type 2 diabetes and mortality) are subject to a range of mechanisms that result in missing data, including: disenrollment from a health plan; that a patient does not have a relevant health encounter at a time point of interest; and, that a measurement was not taken even though standards of care would indicate that it should have been \citep{haneuse2016general, peskoe2021adjusting}.

When it comes to perform analyzes based on data that are incomplete, however, analysts must engage in two related tasks: (i) clarifying and invoking assumptions necessary for valid estimation and inference; and, (ii) employing some statistical methodology that acknowledges the missingness, as appropriate. Regarding the former, it is typical that the missing-at-random (MAR) assumption is invoked \citep{little2019statistical}; intuitively, MAR corresponds to probability of missingness depending only on observed data. In some settings, however, MAR may not hold, in that missingness depends, in part at least, on quantities that are unobserved. This scenario is generally referred to as missing-not-at-random (MNAR) or informative missingness. Unfortunately, in the absence of additional information, MNAR cannot be empirically verified. In turn, much of the literature has focused on sensitivity analysis methods \citep[e.g.,][]{robins2000sensitivity} and estimating bounds on the parameters of interest \citep[e.g.,][]{manski1990nonparametric}. An alternative approach, which we pursue in this work, is to engage in additional data collection efforts, specifically to (re)contact a subset of subjects with missing data to obtain complete data. With this data in-hand, otherwise untestable assumptions can be evaluated (at least partially) and/or incorporated into analyzes. Such a strategy, sometimes referred to as \textit{double-sampling} is well-established as a means to mitigate confounding bias \citep{breslow1996statistics, schildcrout2020two}, bias due to measurement error \citep{lynn2012impact, bound2001measurement} and ecological bias \citep{haneuse2007hierarchical, haneuse2008combination}. However, with few exceptions \citep{frangakis2001addressing, an2009need, qian2020deductive, coppock2017combining, levis2023double}, double-sampling has not been considered routinely as a means to mitigate bias due to potentially MNAR data. 

This paper proposes using double-sampling to mitigate bias due to potentially informative missing data for a continuous outcome when interest lies in estimating contrasts in a class of weighted quantile treatment effects (WQTEs). These estimands, introduced by \cite{sun2021causal}, facilitate comparisons of potential outcomes over various selected target populations, such as among the treated. 
We first put forth conditions sufficient for identifying WQTEs in observational settings with missing outcomes, when the double-sampling design is pursued. Based on this, we propose an IPW approach to estimate WQTEs, labelled IPW-WQTE. We derive the asymptotic distribution of the proposed estimator with estimated propensity scores and missing mechanism at any fixed quantile (pointwise inference), and establish uniform inference of the entire IPW-WQTEs process over a compact subset of $(0,1)$. Finally, we propose a bootstrap resampling method that can be used for either pointwise or uniform inference. The proofs of the main theorems provided in the Supplementary Material. The performance of the proposed approach and estimator is evaluated via a simulation study. The proposed method is then illustrated with data from an EHR-based study examining the relative effects of two bariatric surgery procedures on BMI loss at 3 years post-surgery.

\section{Weighted quantile treatment effects}

In this section, we introduce notation, then briefly review the definition of WQTEs, as well as identification and estimation details in the absence of missing data.

\subsection{Notation}
 Let $Y\in$ denote the continuous outcome of interest and $Z\in\mathcal{Z}=\{0,1\}$ a binary treatment indicator (e.g., $0/1$=control/treatment). The methods in this paper are derived under binary treatment for simplicity, but can be generalized to categorical and continuous treatment using the generalized propensity score \citep[e.g.,][]{imbens2000role, sun2021causal}. Let $X\in\mathcal{X}\subset\mathbbm{R}^p$ denote a $p$-dimensional covariates vector sufficient to control for confounding. We refer to $(Y, Z, X)$ as the `complete data' from a joint distribution $P_c$ used to derive the marginal cumulative distribution function $F_Y(y)=\textrm{pr}(Y\leq y)$, the conditional cumulative distribution $F_{Y\mid Z, X}(y)$, and density function $f_{Y\mid Z, X}(y)$. Let $\mathcal{Y}_{z,x}$ denote the support of the conditional distribution of $Y$ given $Z=z$ and $X=x$. Let $Y(z)$ denote the potential outcome under treatment $Z=z$, and $F_{Y(z)}(y)$ and $F_{Y(z)\mid X}(y)$ denote the corresponding marginal and conditional cumulative distribution functions, respectively, with densities $f_{Y(z)}(y)$ and $f_{Y(z)\mid X}(y)$. Let $\mathbbm{P}_n(U) = n^{-1}\sum_{i=1}^n U_i$ denote the empirical mean of a generic random variable $U$.

\subsection{Weighted quantile treatment effects}

Define the $g$-weighted cumulative distribution function for $Y(z)$,
\begin{equation} \label{eq:wqte2}
    F^g_{Y(z)}(y) = \frac{E\left\{g(X)F_{Y(z)\mid X}(y\mid X)\right\}}{E\left\{g(X)\right\}},\quad z\in\{0, 1\},
\end{equation}
where $g$ is a user-specified function of $X$ for which there exists $c_g>0$ such that $0<g(x)<c_g$ for all $x \in \mathcal{X}$. We then define the $\tau$th WTQE as:
\begin{align}\label{eq:wqte1}
   \beta^g(\tau)\ =\ Q^g_{Y(1)}(\tau) - Q^g_{Y(0)}(\tau),\quad \tau\in(0,1).
\end{align}
\noindent where $Q^g_{Y(z)}(\cdot)$ is the quantile function of $F^g_{Y(z)}(y)$. Note, the introduction of $g(\cdot)$ permits the specification of alternative target populations to which the intent is to generalize. For example, the population QTE \citep{koenker2005quantile} is a special case of the WQTE for $g(x)=1$. As another example, if $g(x)$ equals the propensity score $e(x)= \textrm{pr}(Z = 1\mid X=x)$, the estimand in equation~\eqref{eq:wqte1} corresponds to the $\tau$th QTE among the treated, a QTE analogue to the average treatment effect among the treated. See \cite{sun2021causal} for more details.

\subsection{Identification and estimation given complete data} \label{sec:WQTE:iden}

Identification of $\beta^g(\tau)$ is made possible under the following standard assumptions:
\begin{enumerate}
	\item Stable unit treatment value assumption (SUTVA): For each unit, the potential outcomes are unrelated to the treatment status of other units (no interference), and there are no different forms or versions of each treatment level which could lead to different potential outcomes (consistency).
	\item No unmeasured confounding: $\{Y(0), Y(1)\}\perp Z \mid X$.
	\item Positivity: For some $c_e>0$, $c_e < e(x) < 1-c_e$ for all $x \in \mathcal{X}$.
\end{enumerate}

Focusing on the binary treatment setting, let $Q^g_{Y(0)}(\tau)\eqqcolon\beta^g_0(\tau)$ so that, by the definition given by expression (\ref{eq:wqte1}), $Q^g_{Y(1)}(\tau) = \beta^g_0(\tau)+\beta^g(\tau)$. \cite{sun2021causal} showed that under the three above assumptions, $Q_{Y(z)}^g(\tau)$ is identified by the following population-level estimating equation: for $z \in \{0,1\}$, 
\[0 = E\left(g(X)\frac{\mathbbm{1}\left\{Z = z\right\}}{e(X)^z (1 - e(X))^{1 - z}}\left[\mathbbm{1}\{Y \leq Q_{Y(z)}^g(\tau)\}-\tau\right]\right).\]
These estimating equations motivate the following estimator: given a sample of independent and identically distributed draws from the joint distribution $P_c$, 
\begin{equation}\label{eq:EE_full}
	\mathbbm{P}_n\left( W^g(Z,X)\begin{pmatrix*}
	    1\\[-5pt] Z
	\end{pmatrix*}\left[\mathbbm{1}\{Y<\beta^g_0(\tau)+\beta^g(\tau)Z\}-\tau\right]\right) = 0,\quad \tau\in(0,1),
\end{equation}
with $W^g(Z_i,X_i)=g(X_i)Z_i/e(X_i)+g(X_i)(1-Z_i)/\{1- e(X_i)\}$.

\section{The proposed framework}\label{sec:proposed}
\subsection{Double-sampling}

For settings where $Y$ is potentially missing, let $R\in\{0,1\}$ denote the corresponding observance indicator so that $R=0/1$ if $Y$ is missing/observed. In this context, MAR states, intuitively, that the probability of observing $Y$ only depends on the fully observed variables. Formally, one can write MAR as: $R \perp Y \mid Z, X$. From this formal definition, MNAR may be seen to result when missingness depends on the underlying value of the outcome, either directly or through some (as-yet unspecified) other factor(s). In the bariatric surgery context, for example, the missingness process may be MNAR if patients who do not respond to surgery (i.e., fail to lose weight and/or maintain weight loss) are systematically more likely to have complete data---even after accounting for observed covariates and treatment status---perhaps because they are more likely to continue engaging with the health care system.

In cases where MAR is uncertain, we consider collecting the otherwise missing data on a sub-sample, a process termed `double-sampling' by \cite{frangakis2001addressing}. Let $S\in\{0,1\}$ be an indicator of double-sampling, indicating whether an individual with initially missing outcome data is selected for follow-up and complete data are collected. By design, $S=1$ only if $R=0$ and $S\equiv S(1-R)$. We define the double-sampling probability $\eta(z, x):=\textrm{pr}(S=1\mid Z = z, X = x, R=0)$ for $z\in\mathcal{Z}, x\in \mathcal{X}$. Additionally, the process by which these otherwise missing data are ascertained will be context specific and may involve active participation of study subjects (e.g., response to a survey or phone call) or no participation (e.g., manual chart review); see \cite{levis2023double} for additional details. Under this scheme, the `observed data' are $n$ i.i.d. copies of $\{R, S, (R+S)Y, Z, X\}$ with corresponding joint distribution $P_o$. 

\subsection{Assumptions}\label{subsec:assump}

Here, we outline a set of assumptions used to establish identification of the effects of interest and the asymptotic properties of the estimator proposed in Section \ref{subsec:estimator}. Prior to doing so, we define $D_1(\tau) =E\left[g(X)f_{Y(1)\mid X}\{\beta^{g*}_0(\tau)+\beta^{g*}(\tau)\}\right]$ and $D_0(\tau)=E\left[g(X)f_{Y(0)\mid X}\{\beta^{g*}_0(\tau)\}\right]$, where  $\tau\in (0, 1)$ and $\{\beta^{g*}_0(\tau), \beta^{g*}(\tau)\}$ denote the true values of $\{\beta^g_0(\tau), \beta^g(\tau)\}$.

\begin{assumption}\label{assump:S_indep}
	Non-informative double-sampling: $S \perp Y \mid Z, X, R=0$
\end{assumption}

\begin{assumption}\label{assump:S_positivity}
	Positivity of double-sampling probabilities:  
 For some $c_{\eta}>0$, $c_{\eta}<\eta(z, x)<1-c_{\eta}$ for $z\in\mathcal{Z}, x\in \mathcal{X}$.
\end{assumption} 

\begin{assumption}\label{assump:F&f}
For all $\tau\in(0,1)$, $z\in\mathcal{Z}$, the distribution function $F_{Y\mid Z = z,X}$ is differentiable at $\beta^{g*}_0(\tau)+\beta^{g*}(\tau)z$ with density $f_{Y\mid Z = z,X}$. 
\end{assumption}

\begin{assumption}\label{assump:Dmat}
The derivative $y\mapsto f_{Y\mid Z, X}(y)$ is continuous and bounded in absolute value from above uniformly over $y\in\mathcal{Y}_{z,x}$ and $X,Z$.
\end{assumption}

\begin{assumption}\label{assump:derivative_ps}
	Let $\eta(Z,X;\kappa)$ and $e(X;\gamma)$ denote user-specified parametric models for the double-sampling probabilities and the propensity scores, respectively. We have that $\eta(Z,X;\kappa)$ and $e(X;\gamma)$ are continuously differentiable at true parameters $\kappa^*$ and $\gamma^*$, respectively, with the derivatives being bounded uniformly in $X,Z$.
\end{assumption}

Assumptions \ref{assump:S_indep}, \ref{assump:S_positivity} relate directly to the double-sampling scheme, ensuring the identification of the full data distribution from the observed data. Specifically, Assumption~\ref{assump:S_indep} states that whether or not an individual is double-sampled is independent of the initially unobserved outcome, given the initially observed data. Assumption~\ref{assump:S_positivity} requires that among individuals with initially missing data, the conditional probability of being double-sampled cannot be 0 or 1 in any subgroup defined by $(Z, X)$. Note, Assumption~\ref{assump:S_indep} can be viewed as an MAR assumption for the second phase of data collection. In practice, depending on the second phase sampling method, discrepancies may arise between intended sampling probabilities controlled by researchers and actual data collection methods (e.g., does an individual fill out the survey?). Thus the follow-up sample may or may not have complete data as planned. For example, in a chart review without individuals' direct engagement, it is reasonable to anticipate that those sub-sampled will provide complete data, supporting the plausibility of Assumption~\ref{assump:S_indep}. In other settings, however, such as telephone surveys, some initially selected individuals may decline to participate, leading to an additional layer of non-response \citep{koffman2021investigating}. In such cases, the double-sampling probabilities will have to be estimated, and the plausibility of Assumption~\ref{assump:S_indep} considered in light of this additional missingness.

Assumptions~\ref{assump:F&f}, \ref{assump:Dmat} are standard in quantile-based analyzes~\citep{koenker2005quantile} to control the asymptotic behaviour of the proposed estimator. Assumption~\ref{assump:F&f} ensures that the expectation of the estimating equation $\phi_{\beta}$ in \eqref{eq:phi_theta} is differentiable, allowing the use of the mean value theorem in the proof. Notably, Assumption~\ref{assump:F&f} can be relaxed to one-sided derivatives, enabling the application of a one-sided version of the Mean Value Theorem. Assumption \ref{assump:Dmat} requires that the conditional density function be bounded ensuring finite variance and the use of the mean value theorem. Assumption \ref{assump:derivative_ps} clarifies parametric specifications for the $\eta(Z,X)$ and $e(X)$ when require estimation, acting as regularity conditions for the estimating functions $\phi_{\kappa}$ and $\phi_{\gamma}$ in Section~\ref{subsec:estimator}. Implicit to the assumption is that the class of models adopted for each is sufficiently broad as to contain the true underlying mechanism. Practically, consideration of this assumption will be made in the context of whether and how flexibility is introduced into models specifications, such as using splines. While, arguably, a focus on parametric models may be restrictive, it likely aligns with how most related clinical and public health research is conducted, and so represents an important case to consider. Nevertheless, possible extensions to nonparametric models are discussed in Section~\ref{sec:discussion}.

\subsection{An inverse-probability weighted estimator}\label{subsec:estimator}

In this work, we consider IPW-based estimators of population WQTEs (IPW-WQTEs). Specifically, supposing first that the propensity score and double sampling probabilities are known, we consider for any $\tau \in (0,1)$ the estimator defined as approximately solving:
\begin{equation}\label{eq:DS_EE}
    \mathbbm{P}_n\left(\phi_{\beta^g_0(\tau), \beta^g(\tau)}\right)\ =\ \mathbbm{P}_n\left(\left\{R+\frac{S(1-R)}{\eta(Z, X)}\right\}W^g(Z,X)\begin{pmatrix*}
	    1\\[-5pt] Z
	\end{pmatrix*}\left[\mathbbm{1}\{Y<\beta^g_0(\tau)+\beta^g(\tau)Z\}-\tau\right]\right)\ =\ 0
\end{equation}
Expression (\ref{eq:DS_EE}) is derived by splitting the estimating equation~\eqref{eq:EE_full} into two parts: one for the originally observed individuals (those with $R$ = 1) and another for the double-sampled individuals (those with $S$ = 1), appropriately weighted according to the observed data (see Section S.1. of the Supplementary Materials). By `approximately' we mean that $\|\mathbbm{P}_n(\phi_{\beta^g_0(\tau), \beta^g(\tau)})\|=o_P(n^{-1/2})$ and the solution to \eqref{eq:DS_EE} $\{\widehat{\beta}_0^g(\tau),\widehat{\beta}^g(\tau)\}\overset{p}{\to} \{\beta^{g*}_0(\tau), \beta^{g*}(\tau)\}$ \citep{van2000asymptotic}. Furthermore, under Assumptions~\ref{assump:S_indep} and \ref{assump:S_positivity}, by the first order condition, we have
\begin{align}
    E\left(\left\{R+\frac{S(1-R)}{\eta(Z, X)}\right\}W^g(Z,X)\begin{pmatrix*}
	    1\\[-5pt] Z
	\end{pmatrix*}\left[\mathbbm{1}\{Y<\beta^{g*}_0(\tau)+\beta^{g*}(\tau)Z\}\right]-\tau\right) = 0
\end{align}
at the true values $\{\beta^{g*}_0(\tau),\beta^{g*}(\tau)\}$, which motivates our approach.

In practice, as discussed in Section~\ref{subsec:assump}, 
the true propensity score and, possibly, the double-sampling probability, will need to be estimated. Following Assumption~\ref{assump:derivative_ps}, suppose the parameters indexing the models $\eta(Z,X;\kappa)$ and $e(X;\gamma)$ are estimated in-sample (i.e., using the data we have), obtained as the solutions to estimating equations based on estimating functions $\phi_{\kappa}$ and $\phi_{\gamma}$, respectively. For example, $\eta(Z,X;\kappa)$ and $e(X;\gamma)$ can be estimated using logistic regressions, with $\phi_{\kappa}$ and $\phi_{\gamma}$ representing the corresponding score functions. An example of such a score function is $\phi_{\gamma}=\sum_{i=1}^n[y_i-\{1+\exp(-\boldsymbol{x}_i^\top\gamma)\}^{-1}]\boldsymbol{x}_i^\top$.  Subsequently, $\widehat{\kappa}$ and $\widehat{\gamma}$ denote the maximum likelihood estimators for these parameters. 

Plugging in the estimated propensity scores and double-sampling probabilities into expression \eqref{eq:DS_EE} yields the proposed estimator as approximately solving an equation:
\begin{equation}\label{eq:DS_EE_est}
	\mathbbm{P}_n\left(\left\{R+\frac{S(1-R)}{\eta(Z, X;\widehat{\kappa})}\right\}{W^g}(Z,X;\widehat{\gamma})\begin{pmatrix*}
	    1\\[-5pt] Z
	\end{pmatrix*}\left[\mathbbm{1}\{Y<\beta^g_0(\tau)+\beta^g(\tau)Z\}-\tau\right]\right) = 0.  
\end{equation}
Equivalently, letting $\theta = \{\kappa, \gamma, \beta^g_0(\tau), \beta^g(\tau)\}$ and $\Phi_{\theta} = (\phi_{\kappa},\phi_{\gamma},\phi_{\beta})$ with 
\begin{equation}\label{eq:phi_theta}
    \phi_{\beta} = \left\{R+\frac{S(1-R)}{\eta(Z, X;\kappa)}\right\}W^g(Z,X;\gamma)\begin{pmatrix*}
	    1\\[-5pt] Z
	\end{pmatrix*}\left[\mathbbm{1}\{Y<\beta^g_0(\tau)+\beta^g(\tau)Z\}-\tau\right],
\end{equation}
the proposed IPW-WQTE estimator $\widehat{\beta}^g(\tau)$ in \eqref{eq:DS_EE_est} approximately solves the equation $\mathbbm{P}_n(\Phi_{\theta})=0$. Again, by `approximately' we mean that $\|\mathbbm{P}_n(\Phi_{\widehat\theta})\|=o_P(n^{-1/2})$ and $\widehat\theta\overset{p}{\to} \theta^*$ \citep{van2000asymptotic}. In the following two sections, we establish asymptotic properties of the proposed estimator under the scenarios where interest lies in some specific quantile (e.g., the median or some high quantile level such as the 0.9 quantile) and where interest lies in a range of quantiles over some compact subset $\mathcal{T}$ of $(0,1)$ (e.g., the lower tail such as $[0.05, 0.2]$).

\subsection{Asymptotic theory for fixed quantiles}\label{subsec:asymp_fix}
Suppose interest lies in estimating the causal WQTE for some fixed quantile level $\tau\in(0,1)$. The following result establishes pointwise properties of $\widehat{\beta}^g(\tau)$: 

\begin{theorem}\label{thm:asymp}
	Under Assumptions~\ref{assump:S_indep}--\ref{assump:F&f} and \ref{assump:derivative_ps}, and assuming $D_1(\tau), D_0(\tau)\in (0,\infty)$ for any fixed $\tau\in(0,1)$, then $\widehat{\beta}^g(\tau)$ is consistent to $\beta^{g*}(\tau)$, that is, $\widehat{\beta}^g(\tau)\overset{p}{\to}\beta^{g*}(\tau)$ as $n\rightarrow\infty$, and 
\begin{equation*}
    n^{1/2}\{\widehat{\beta}^g(\tau)-\beta^{g*}(\tau)\} =  \boldsymbol{e}_2^\top\Sigma^{-1}(\tau)\mathbbm{G}_n(\psi_{\theta^*})+ o_p(1),
\end{equation*}
where $\boldsymbol{e}_2=(0,1)^\top$, 
\begin{align*}
&\Sigma(\tau) = \begin{pmatrix*}[l]
        D_1(\tau)+D_0(\tau)& D_1(\tau)\\
        D_1(\tau)& D_1(\tau)\\
    \end{pmatrix*},\\
&\psi_{\theta} = V_{\kappa,3}V^{-1}_{\kappa,1}\phi_{\kappa}+V_{\gamma,3}V^{-1}_{\gamma,2}\phi_{\gamma} -\phi_{\beta},
\end{align*}
$V_{\kappa,1}$ and $V_{\gamma, 2}$ respectively are the derivative of $E(\phi_{\kappa})$ and $E(\phi_{\gamma})$, and $V_{\kappa,3}$ and $V_{\gamma,3}$ sequentially are the derivatives of $E(\phi_{\beta})$ at $\kappa$ and $\gamma$, and $\mathbbm{G}_nf=\sqrt{n}(\mathbbm{P}_n-P)f$.
\end{theorem}
While complete details are given in Section S.2. of the Supplementary Materials, briefly, the proof of Theorem~\ref{thm:asymp}, which includes proof of consistency of $\widehat{\beta}^g(\tau)$, first involves showing that the class $\{\Phi_{\theta}: ||\theta-\theta^*||<\delta\}$ is $P$-Donsker for some $\delta>0$ with finite envelope function. Theorem 19.26 of \cite{van2000asymptotic} can then be invoked directly. Note that the vector $\boldsymbol{e}_2$ is used to derive the linear representation for the proposed IPW-WQTE estimator $\widehat{\beta}^g(\tau)$, which is one of the two coefficients in equation~\eqref{eq:DS_EE_est}. Crucially, Theorem~\ref{thm:asymp} establishes that the proposed IPW-WQTE estimator, $\widehat{\beta}^g(\tau)$, can be written in a form of linear representation as a normalized sum of i.i.d. random variables, with a negligible remainder, so that, in turn, the central limit theorem can be applied. Remarkably, the variance arising from estimating the nuisance parameters $\kappa$, $\gamma$ and $\beta_0^{g}(\tau)$ is encompassed withing $\psi_{\theta}$.

\begin{corollary}\label{cor:asymp_norm}
	Assume the assumptions in Theorem~\ref{thm:asymp} hold. For any fixed $\tau\in(0,1)$, $n^{1/2}\{\widehat{\beta}^g(\tau)-\beta^{g*}(\tau)\}$ converges in distribution to $N(0,\boldsymbol{e}_2^\top\Sigma^{-1}cov(\psi_{\theta^*})\Sigma^{-1}\boldsymbol{e}_2)$ as $n\rightarrow\infty$.
\end{corollary}

\subsection{Asymptotic theory for the entire quantile process}\label{subsec:asymp_proc}

For settings where interest lies in estimating causal WQTEs across a range of quantiles, we extend the asymptotic theory regarding $\widehat{\beta}^g(\tau)$ presented above to the entire process $\{\widehat{\beta}^g(\tau): \tau\in\mathcal{T}\}$ with $\mathcal{T}$ some compact subset of $(0,1)$. For an arbitrary set $\mathcal{T}$, let $\ell^{\infty}(\mathcal{T})$ denote the space of all bounded functions from set $\mathcal{T}$ to $\mathbbm{R}$ equipped with the uniform norm $\|f\|_{\mathcal{T}}=\sup_{\tau\in\mathcal{T}}|f(\tau)|$. Specifically, the following Theorem establishes key asymptotic properties of the process $M_n = \left[n^{1/2}\{\widehat{\beta}^g(\tau)-\beta^{g*}(\tau)\}: \tau\in \mathcal{T}\right]$:
\begin{theorem}\label{thm:unif_asymp}
Under Assumptions~\ref{assump:S_indep}--\ref{assump:derivative_ps}, for any $\tau_1, \tau_2 \in \mathcal{T}$, $\widehat{\beta}^g(\tau)$ is uniformly consistent to $\beta^{g*}(\tau)$, that is, $\sup_{\tau\in\mathcal{T}}\|\widehat{\beta}^g(\tau)-\beta^{g*}(\tau)\|\overset{p}{\to} 0$ as $n\rightarrow\infty$, and the sequence of maps $M_n$ convergences weakly to a measurable tight process $M_{\infty}$ in $\ell^{\infty}(\mathcal{T})$ as $n\rightarrow\infty$, where $M_{\infty}$ is a centred Gaussian process with covariance 
\begin{equation*}
    cov\{M_{\infty}(\tau_1)M_{\infty}(\tau_2)\}=\boldsymbol{e}^\top_2\Sigma^{-1}(\tau_1)\Omega(\tau_1,\tau_2)\Sigma^{-1}(\tau_2)\boldsymbol{e}_2
\end{equation*}
where $\Omega(\tau_1,\tau_2)$ is the asymptotic covariance between $\mathbbm{G}_n(\psi_{\theta^*})$ at $\tau_1$ and $\tau_2$.
\end{theorem}The proof of Theorem \ref{thm:unif_asymp} can be found in Section S.2 of the Supplementary Materials.

\subsection{Variance estimation via the bootstrap}\label{subsec:bootstrap}

Practically, variance estimation for both the fixed quantile and uniform inference settings will likely be challenging. In each case, the asymptotic variance has a complex form, and estimating it requires estimates of the potential outcome conditional densities (i.e., $f_{Y(1)\mid X}$ and $f_{Y(0)\mid X}$). Moreover, estimating these will be complex if, as is often the case, $X$ includes many covariates of different types. To simplify the variance calculation, we propose alternatives based on the bootstrap. For pointwise inference regarding some fixed quantile level, we note that any standard bootstrap approach can be used. For example, one could proceed by performing the same estimation procedure on a bootstrap replicate the same way as in the original data set. To construct uniform confidence bands and perform uniform inference on the entire IPW-WQTE process, we present the following theorem that justifies a procedure inspired by the gradient bootstrap resampling method proposed by \cite{belloni2019conditional}. Our procedure differs, however, in that the coefficient process is derived from a series of weighted quantile regressions. Specifically, we define a gradient bootstrap estimator of $\beta^{g}(\tau)$ as a solution to the perturbed problem:
\begin{align*}
\widehat{\beta}^{g^*}(\tau)\in \mathrm{argmin}_{\beta^g_0(\tau),\beta^g(\tau)}\mathbbm{P}_n&\Bigg[\left\{R_i+\frac{S_i(1-R_i)}{\eta(Z_i, X_i;\widetilde{\kappa})}\right\}{W^g}(Z_i,X_i;\widetilde{\gamma})\rho_{\tau}\{Y_i-\beta^g_0(\tau)-\beta^g(\tau) Z_i\}\\
&\quad-\mathbbm{U}^*(\beta^g_0(\tau),\beta^g(\tau);\tau)/\sqrt{n}\Bigg],
\end{align*}
where $\rho_{\tau}(u)=\tau\{\tau-\mathbbm{1}(u\leq \tau)\}$ is the check function, $D=\{(R_i,S_i,(R_i+S_i)Y_i,Z_i,X_i): 1\leq i\leq n\}$ is the original data, $\widetilde{\kappa}$ and $ \widetilde{\gamma}$ are the estimated nuisance parameters using the bootstrap sample $D^b=\{(R^b_i, S^b_i, (R^b_i+S^b_i)Y^b_i, Z^b_i, X^b_i): 1\leq i\leq n\}$, and 
\begin{align*}
    \mathbbm{U}^*(\beta^g_0(\tau),\beta(\tau);\tau)=\ \frac{1}{\sqrt{n}}\sum_{i=1}^n \left\{R_i+\frac{S_i(1-R_i)}{\eta(Z_i, X_i;\widetilde{\kappa}})\right\}{W^g}(Z_i,X_i;\widetilde{\gamma}){\{\beta^g_0(\tau)+\beta^g(\tau)Z_i\}}\{\tau-\mathbbm{1}(U^b_i\leq\tau)\},
\end{align*}
where for each $D^b$, $\{U^b_i: 1\leq i\leq n\}$ is a sequence of i.i.d. $U(0,1)$ random variables. The following theorem justifies the validity of this gradient bootstrap estimator of $\beta^g(\tau)$ and thereafter its use to construct data-driven uniform confidence bands for $\beta^g(\tau),\tau\in \mathcal{T}$.

\begin{theorem}\label{thm:bootstrap}
Suppose that the conditions in Theorem~\ref{thm:unif_asymp} hold. Then the sequence of maps $\widehat{M}_n=\left[n^{1/2}\{\widehat{\beta}^g(\tau)-\widehat{\beta}^{g*}(\tau)\}: \tau\in \mathcal{T}\right]$ convergences weakly to a measurable tight process $M_{\infty}$ in $\ell^{\infty}(\mathcal{T})$ defined in Theorem \ref{thm:unif_asymp}.
\end{theorem}
The proof of Theorem \ref{thm:bootstrap} can be found in Section S.2 of the Supplementary Materials. As mentioned above, Theorem \ref{thm:bootstrap} allows us to compute (asymptotically) uniformly valid confidence bands for the process $\beta^g(\tau),\tau\in \mathcal{T}$ by simulating the process $ \mathbbm{U}^*(\beta^g_0(\tau),\beta(\tau);\tau)$. The algorithm is provided in Section S.3 of the Supplementary Materials. 

In practice, if researchers do not have a predefined quantile level of interest and aim to estimate WQTEs across multiple quantiles, Theorems~\ref{thm:unif_asymp} and \ref{thm:bootstrap} ensure asymptotically valid inference over any compact set in $(0,1)$. Importantly, the uniform confidence bands facilitate global inference across a compact set $\mathcal{T}$ rather than just a specific point; any null hypothesis that lies outside of the uniform confidence bands, even at a single quantile, can be rejected at a pre-specified significant level. For example, if the band fails to contain 0 at any quantile in $\mathcal{T}$, we can reject the null hypothesis that there is no causal effect across $\mathcal{T}$.

\section{Simulation}\label{sec:simulation}

We conducted a series of simulation studies to investigate the properties of the methods proposed in Section~\ref{sec:proposed}. Throughout, for simplicity, we focus on estimation of the population QTE, that is $\beta^g(\tau)$ given by expression (\ref{eq:wqte1}) with $g(X)=1$.

\subsection{Data generation}

For each simulation scenario we generated 10,000 datasets, each with sample size $n=10,000$. For each observation within each dataset, we generated two covariates: $X_1\sim U(0, 1)$ and $X_2\sim U(0,2)$. Letting $X \equiv (X_1, X_2)$, a binary `treatment' variable was generated as $Z\mid X\sim \mbox{Bernoulli}\{e(X)\}$, with $e(X)=\{1+\exp(-0.5+0.5X_1+0.5X_2)\}^{-1}$. Finally, the outcome $Y$ was generated using the linear model $Y=1+Z+X_1+X_2+(1+\rho Z)\varepsilon$, where $\varepsilon$ is the error term and $\rho$ is a constant controlling the heteroscedasticity of QTEs across quantiles $\tau\in (0,1)$. We used a Pareto distribution for $\varepsilon$, inducing a right-skewed marginal distribution of $Y$ (Figure S.1). We considered two cases for $\rho$: (i) $\rho=0$, resulting in homogeneous QTEs of 1; and (ii) $\rho=1.5$, resulting in heterogeneous QTEs from 2.67 to 3.10 (Figure S.2).

Missingness was subsequently introduced to reflect the primary scenario that the proposed methods were designed to address, that being missing outcomes subject to MNAR. Towards this, we generated an initial observance indicator via, $R\mid Y \sim \mbox{Bernoulli}\{\pi(Y)\}$: for the homogeneous QTE scenario, we set $\pi(Y)=(1+\exp(-1-4.3Y+Y^2))^{-1}$; for the heterogeneous QTEs scenario, we set $\pi(Y)=(1+\exp(-1-3.8Y+Y^{1.8}))^{-1}$. Note, both of these specifications resulted in marginal missingness probabilities of approximately 35\%.

Finally, we considered double-sampling those with $R=0$ within strata defined by $(Z, X_1, X_2)$, where $X_1$ and $X_2$ were dichotomized at 0.5 and 1, respectively. We pre-defined the stratum-specific sample size varying from 0 and 240 in each of the eight strata (see Table S.1 and S.2 in Section S.4.1 of the Supplementary Materials). To ensure the non-informative double-sampling assumption, a stratum-specific sample was randomly `selected' in each stratum. This procedure `selected' a total of double-sampled 800 individuals with initially missing outcomes. Overall these choices resulted in double-sampling of approximately 22\% of individuals with original missing data.

\subsection{Analysis}\label{subsec:analysis}

We estimate the population QTE, $\beta^g(\tau)$ with $g(X)=1$, for $\tau$ from 0.1 to 0.9 with increments of 0.1. Specifically, we considered five estimators: (I) an estimator based on the full data, with $e(X)$ estimated via maximum likelihood; (II) an estimator based on the initial incomplete data alone (excluding double-sampled data), with $e(X)$ estimated via maximum likelihood; (III) an estimator based on the initial incomplete data coupled with the double-sampled data, with $e(X)$ set at their true values and $\eta(Z,X)$ estimated via maximum likelihood; (IV) an estimator based on the initial incomplete data coupled with the double-sampled data, with both $e(X)$ and $\eta(Z,X)$ estimated via maximum likelihood; and, (V) an estimator based on the initial incomplete data assuming MAR, with both $e(X)$ and $\pi(Z,X):=pr(R=1\mid Z,X)$ estimated via maximum likelihood. Specifically, estimator (V) approximately solves an equation $\mathbbm{P}_n\left(\frac{R}{\pi(Z, X;\widehat{\alpha})}{W^g}(Z,X;\widehat{\gamma})(1, Z)^\top\left[\mathbbm{1}\{Y<\beta^g_0(\tau)+\beta^g(\tau)Z\}-\tau\right]\right) = 0$. Note, throughout, estimates of $e(X)$ and $\eta(Z,X)$ were based on a correctly-specific logistic regression. Furthermore, since we performed a stratified sampling with fixed sample sizes, and the eight potential strata had random sample sizes (due to the re-generation of the covariate data), there is no single set of `true' values for the double-sampling probabilities. We thus estimated $\eta(Z,X)$ via a saturated logistic regression on the eight strata.

Asymptotic variance calculations for proposed estimators (III) and (IV) were computed assuming the true underlying distribution family. Specifically, for estimator (III), the function $\psi_{\theta}$ from Theorem~\ref{thm:asymp} simplifies to $V_{\kappa,3}V^{-1}_{\kappa,1}\phi_{\kappa}-\phi_{\beta}$. The asymptotic variance was then calculated with the true $e(x)$ and estimated parameters $\{\widehat{\kappa}, \widehat{\beta}_0^g(\tau), \widehat{\beta}^g(\tau)\}$ plugged into $\boldsymbol{e}_2^\top\Sigma^{-1}cov(\psi_{\theta})\Sigma^{-1}\boldsymbol{e}_2$. For estimator (IV), we first estimated parameters $\widehat{\theta} = \{\widehat{\kappa}, \widehat{\gamma}, \widehat{\beta}_0^g(\tau), \widehat{\beta}^g(\tau)\}$, then plugged them into the corresponding asymptotic variance expression of (IV), as presented in Theorem~\ref{thm:asymp} and Corollary~\ref{cor:asymp_norm}. Pointwise bootstrap variance estimates were computed for estimators (III) and (IV) at each $\tau$ considered, and the 95\% confidence bands were constructed using the gradient bootstrap method (Section~\ref{subsec:asymp_proc}). 
Coverage was assessed by determining if the band included the true QTE for each quantile; if all values were included, coverage was successful for the given realization; otherwise, it is considered a failure. The empirical coverage probability is the proportion of realizations with successful coverage of the true parameters.

\subsection{Results}

\begin{figure}
    \centering
    \includegraphics[scale=0.6]{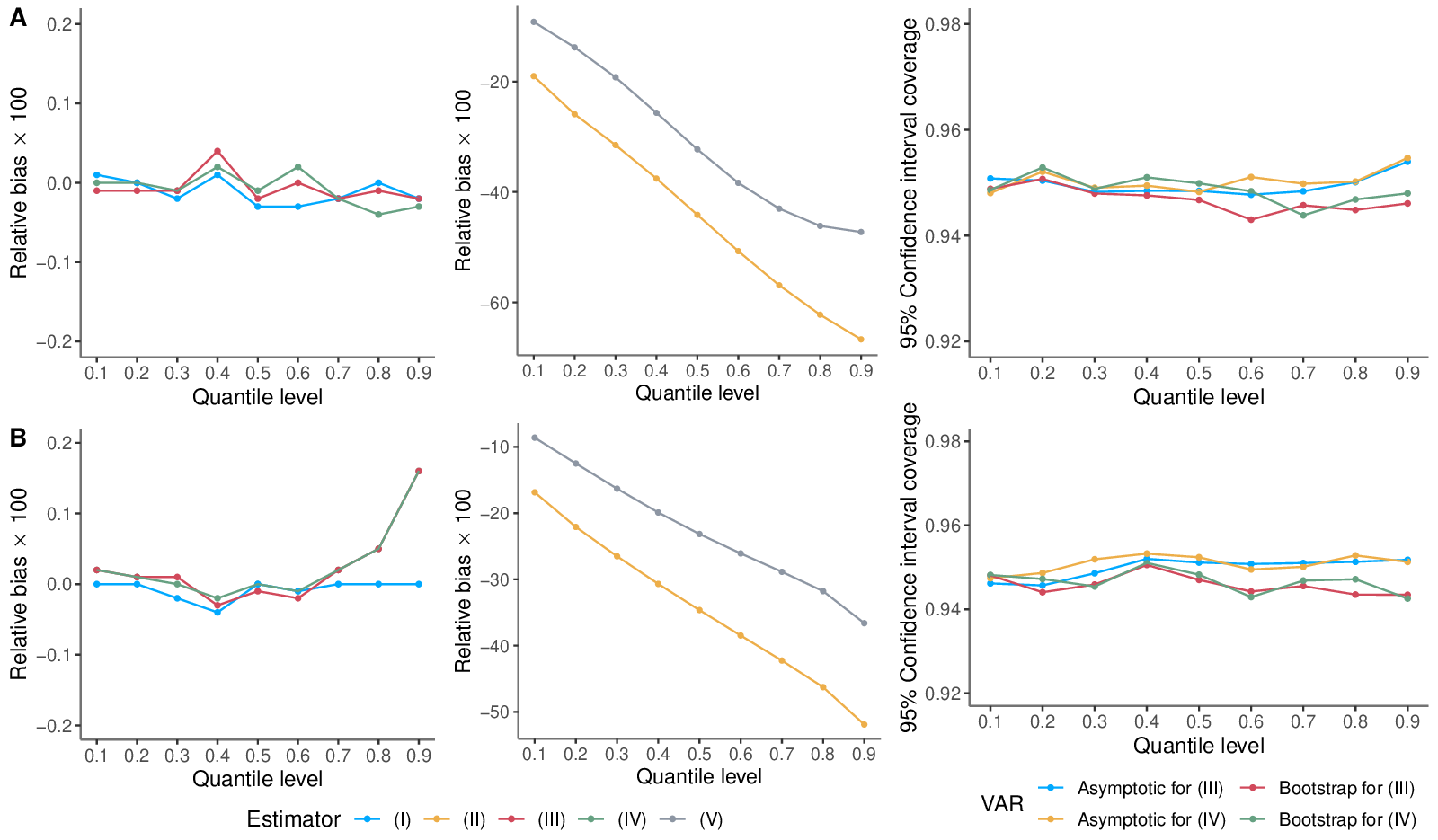}
    \caption{Pointwise relative bias and 95\% confidence interval coverages of QTE estimates across quantile levels. Row (A): Homogeneous QTE setting. Row (B): Heterogeneous QTE setting. Left and middle columns: Relative bias. Right column: Pointwise 95\% confidence interval coverage Estimators (I)--(V) denote the five estimators in Section~\ref{subsec:analysis}. The asymptotic variance and bootstrap variance are defined in Section~\ref{subsec:analysis}. Total 10,000 datasets, each with sample size 10,000. Abbreviation: VAR, variance.}
    \label{fig:sim_QTE}
\end{figure}

Figure~\ref{fig:sim_QTE} presents point-wise relative bias ($\times$ 100) and 95\% confidence interval coverage for the homogeneous and heterogeneous QTE scenarios. Throughout, we see that the proposed estimators (III) and (IV) in Section~\ref{subsec:analysis} have relative bias and confidence interval coverage comparable to the estimator (I) based on the full data. Regarding the 95\% confidence bands for uniform inference, coverage was 98.6\% for estimation of the homogeneous QTE and 98.9\% for estimation of the heterogeneous QTE. Note, this over coverage is consistent with prior reports regarding confidence band estimation in finite samples \citep{belloni2019conditional,song2012bootstrap,kocherginsky2005practical}. Finally, the complete case estimator (II) and estimator (V) assuming MAR are both substantially biased.

\begin{figure}[htbp]
    \centering
    \includegraphics[scale=0.6]{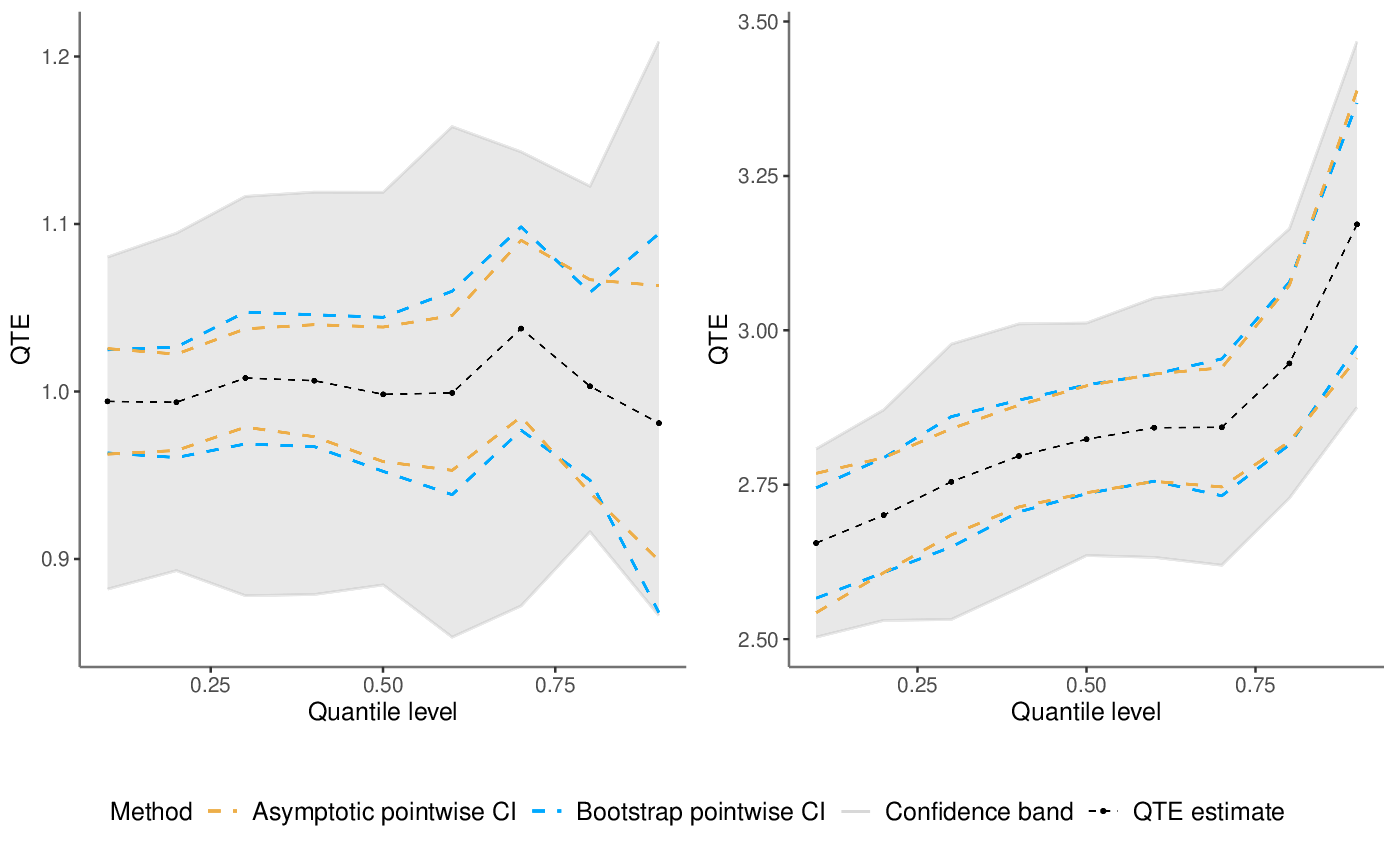}
    \caption{Comparison of pointwise 95\% confidence intervals versus uniform bands for the point QTE estimates of a realization. The left panel is in homogeneous QTE setting. The right panel is in heterogeneous QTE setting. Pointwise confidence interval were estimated using the asymptotic and bootstrap variance, as defined in Section~\ref{subsec:analysis}; the confidence band is outlined in the same section. Abbreviation: CI, confidence interval.}
    \label{fig:cb}
\end{figure}

Figure~\ref{fig:cb} illustrates results for a single dataset, comparing 95\% pointwise confidence intervals (asymptotic and bootstrap-based) with confidence bands in homogeneous and heterogeneous QTE settings. The black dotted line shows the point QTE estimates, the blue and yellow dashed lines show the 95\% pointwise CIs, and the light grey areas show the 95\% uniform confidence bands, which are wider by construction. In the left panel, QTE estimates demonstrate similarity across quantiles; whereas in the right panel, QTE estimates show more beneficial treatment effects with larger $Y$ values. Notably, the heavy-tailed error term $\varepsilon$ widens pointwise CIs on the right due to the density function in the variance formula (Theorem~\ref{thm:asymp}). Thus, the disparity between confidence bands and intervals narrows in the upper right of Figure~\ref{fig:cb}.
Similar patterns show with missingness rates of 20\%, 50\%, and 70\% and smaller samples (2,000 and 5,000) (Section S.4.3 of the Supplementary Materials).

\section{Data Application}

In this section, we illustrate the proposed methods with a comparison of the relative effects of two bariatric surgery procedures on long-term weight loss using EHR data from a broader investigation of long-term outcomes for patients who had undergone bariatric surgery in one of the three Kaiser Permanente health care sites: Washington, Northern California, and Southern California\citep{arterburn2021weight}. In the present study, we aim to compare Roux-en-Y gastric bypass (RYGB) ($Z=1$) versus vertical sleeve gastrectomy (VSG) ($Z=0$) bariatric surgery procedures on percent BMI change at 3 years post-surgery ($Y$). From a larger sample of 30,991 patients for which follow-up outcomes were partially observed, we selected $13,514$ patients who underwent RYGB or VSG between January 2005 and September 2015, with complete weight data at baseline (closest measurement pre-surgery, up to 6 months) and at follow-up (closest measurement within $\pm$ 90 days). See Table~S.3 for a summary of the baseline characteristics of individuals who underwent bariatric surgeries.

For the study sample of $13,514$ patients, we then artificially imposed a MNAR missingness mechanism in the outcome. To accomplish this, we first modelled the probability of observance indicator ($R$) from the original larger population of 30,991 patients using the following baseline covariates ($X$): baseline weight, health care site, year of surgery, age, gender, ethnicity, number of days of health care use in 7-12 month period pre-surgery, number of days hospitalized in pre-surgery year, smoking status, Charlson/Elixhauser comorbidity score, insurance type, clinical statuses for hypertension, coronary artery disease, diabetes, dyslipidemia, retinopathy, neuropathy, mental health disorders, and use of medicines which includes insulin, ACE inhibitors, ARB, statins, other lipid lowering medications, and other antihypertensives. The model for $\textrm{pr}(R=1\mid Z, X)$ was fit using a SuperLearner ensemble with library \{\texttt{SL.glm}, \texttt{SL.ranger}, \texttt{SL.rpart}\} in \texttt{R} version 4.2.1 \citep{polley2017super}.

Subsequently, to impose MNAR, we additionally included a dependence on the outcome $Y$ by augmenting the fitted values $\widetilde{p}$ of $\textrm{pr}(R=1\mid Z, X)$ via: $\widetilde{p}\ \mapsto p$ where $p=\textrm{expit}\{\textrm{logit}(\widetilde{p})+Y^*[2-Z+0.3(\text{gender})+0.3(\text{race})-(\text{diabetes}) -0.4(\text{statin})-0.5(\text{insulin})]
\}$ where ${Y^*}$ is the standardized 3-year BMI change. Finally, we trimmed $p$ using $0.05$ and $0.95$. To generate MNAR outcomes on the sample of 13,514 patients, we sampled $R$ according to a Bernoulli with probability given by the trimmed $p$. The resulting marginal missingness probability was 30\%. We considered double-sampling subsamples of size 1,000 and 1,500 from all patients with initially missing outcomes using simple random sampling, and estimated the double-sampling probability based on the sample proportion. For each of the two resulting data sets, we computed the QTEs at a sequence of quantile levels (i.e., $0.1, 0.3, 0.5, 0.7, 0.9$), and compared the point estimates and 95\% pointwise bootstrap CIs using the proposed method in Section~\ref{sec:proposed}. A benchmark QTE estimate for each quantile level was estimated using the full data set (before artificial imposition of missing outcomes). For comparison, we also estimated QTEs using the complete-case data set ($R=1$). We computed 95\% confidence bands using the data set with 1,500 double-sampled patients and compared to the pointwise bootstrap confidence intervals. Additionally, as the effects of the two bariatric surgeries may be different among patient with diabetes or dyslipidemia diagnosis, we repeated the above analyzes stratifying by diabetes and dyslipidemia diagnosis categories.

\begin{figure}[htbp]
    \centering
    \includegraphics[scale=0.47]{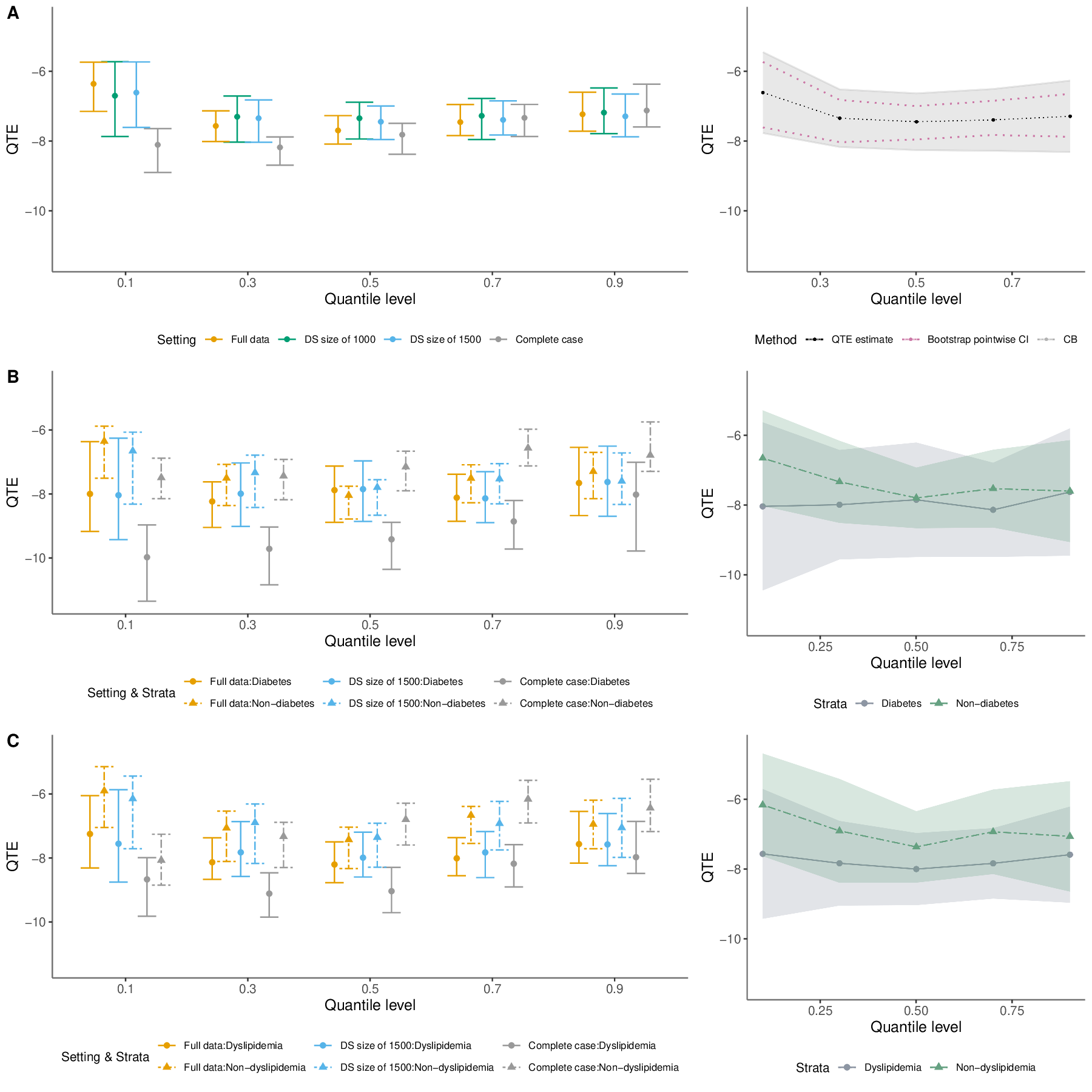}
    \caption{(A) The $n$=13,514 sample. (B) Stratified by diabetes status. (C) Stratified by dyslipidemia diagnosis. Left column: Pointwise QTE estimates and 95\% pointwise bootstrap CIs across quantile levels; point estimates are marked with dots. Right column: Comparison of 95\% pointwise boostrap CIs versus uniform confidence bands; CIs are the dashed lines and CBs are the shaded areas. Abbreviation: DS, double-sample; CI, confidence interval; CB, confidence band.}
    \label{fig:point_qte}
\end{figure}

In Figure~\ref{fig:point_qte}, the left column displays the pointwise QTE estimates and 95\% CIs across quantile levels , and the right column compares pointwise CIs to confidence bands constructed in Section \ref{subsec:bootstrap}. Panels (A) to (C) represent different data subsets: (A) for all 13,514 patients, and (B) and (C) are stratified by diabetes status and dyslipidemia diagnosis, respectively. In panel (A), the confidence band is slightly wider than the pointwise CIs, likely attributable to idiosyncrasies in the real data application. The specific estimated QTE values and 95\% pointwise CIs are listed in Table S.4 in the Supplementary Materials. The results show that complete case estimates are biased compared to the benchmark estimates using the full data, especially at low quantile levels. The proposed IPW-WTQE estimators were   unbiased, with the setting with 1,500 subjects in the second-stage sample showed greater efficiency (i.e. with narrower 95\% CIs) compared to the setting with the smaller subsample. Heterogeneous QTEs were observed across quantile levels as well as at different strata: when compared to VSG, RYGB appears to be associated with greater percent BMI loss at 3 years: (i) at low tails of the percent BMI change distribution, and (ii) among patients with diabetes or dyslipidemia diagnosis at most of the selected quantile levels.

\section{Discussion}\label{sec:discussion}

The work of this paper is motivated, in part, by \cite{koffman2021investigating}, who report on an NIH-funded study that used a telephone survey to investigate missingness in EHR-based weight measurements, relevant to a broader study of long-term outcomes following bariatric surgery. Key to that publication is that it demonstrated the potential for the use of double-sampling in practical settings, specifically as a means to collect information on covariates that are potentially MNAR and learn about missingness mechanisms. The contribution of this paper, then, is to propose a novel approach for estimation and inference for causal weighted QTEs, one that combines the readily-available data in, say, the EHR, with additional data collected on a sub-sample of study units with otherwise missing outc ome information, and a set of proposed identifying assumptions. When estimating contrasts involving the mean of the counterfactual distribution such as the ATE, one may employ $g$-methods that first compute a conditional ATE, conditional on $X$, and then marginalize to recover an estimate of the population causal effect \citep{hernan2020}. However, due to Jensen's inequality, integrating a conditional QTE will not yield the population QTE \citep{sun2021causal}. Motivated by this, we propose a novel inverse-probability weighted estimator as well as a bootstrap-based approach for valid inference, either pointwise for a single quantile or uniformly for a range. The procedure involves first estimating propensity scores and double-sampling probabilities, and then plugging these into an estimating function for the weighted QTE. In a simulation study, we consider settings where the QTEs are either homogeneous or heterogeneous across quantiles, verify theoretical estimation properties in a finite sample setting and demonstrate validity of the proposed bootstrap-based confidence intervals/bands.

Key to the proposed methods is that no assumptions are required regarding the missingness mechanism in the original data. This, arguably, is a critical benefit because mechanisms underlying $R$, especially in the increasingly common setting of using EHR data, are typically outside of the control of the research team and potentially quite complex. Thought experiments on assumptions necessary for progress with the original data (e.g., to consider the potential validity of MAR) can be challenging and inconclusive. Thus, one way of framing the proposed methods is that assumptions regarding $R$ (as would have to be the case if only the original data were available) are replaced with the assumptions regarding the double-sampling indicator $S$; that is, Assumptions \ref{assump:S_indep} and \ref{assump:S_positivity}. In practice, depending on the context and the nature of the additional data to be collected, consideration of these assumptions may be fairly straightforward. For example, suppose the additional data collection involves manual chart review or natural language processing of a sub-sample chosen by the research team~\citep{weiskopf2019towards}, Assumptions \ref{assump:S_indep} and \ref{assump:S_positivity} may hold trivially by design, since no involvement of the study subject is needed. In other settings, however, in particular when study subject participation is needed (e.g., to respond to a survey or phone call), assessment of these assumptions may require more detailed consideration. However, even in these situations, researchers may pursue additional design options that may shed light. For example, they may employ intensive, repeated efforts to contact study subjects and collect the data and/or engage with study subjects who have complete information in the original data to learn about potential recall bias \citep{haneuse2016learning, koffman2021investigating}. EHR studies often encompass vast datasets, yet the ability to reach out to patients typically does not grow with sample size. Given that the double-sampling framework can be strategically planned from the beginning, it presents an opportunity to allocate resources effectively within the budgetary confines. Our ongoing research aims to find an optimal sampling rule that minimizes the semiparametric efficiency bound subject to a budget constraint.

There are numerous opportunities for building on the proposed framework. One such opportunity is to consider settings beyond binary treatment settings, including more than two discrete options as well as continuous treatments (e.g., dose) \citep{sun2021causal}. A second opportunity is to consider alternatives to parametric models for the propensity scores and double-sampling probabilities, such as more flexible machine learning techniques \citep{chernozhukov2018double}. For the latter, an important challenge will be that the rate of convergence for estimation in non-parametric models will affect both the achievable rates of convergence and asymptotic distribution of the weighted QTE estimator. Relatedly, we note that \cite{belloni2019conditional} developed a theory for inferring the unweighted QTE in the absence of missing data, under specific assumptions on the nonparametric models for the quantile regression function. Even in this case without missing data, however, the complete picture of nonparametric estimation of the unweighted QTE through the lens of a higher-order semiparametric theory \citep{robins2008higher} remains open. Therefore, here we keep a detailed study of the nonparametric analogue of the problem we consider for future directions. Finally, it will often be the case that covariates beyond the outcome may be missing (e.g., an important confounder); extending the assumptions and estimator to acknowledge this is an avenue we are currently investigating.


\backmatter


\section*{Acknowledgements}
The authors gratefully acknowledge support from National Institutes of Health grants R-01 DK128150 and R-01 HL166324.\vspace*{-8pt}

\section*{Supplementary Materials}
Web Appendices S.1, S.2, and S.3 referenced in Section 3, Web Appendices S.4, Tables S.1 to S.2, and Figures S.1 to S.8 referenced in Section 4, Web Appendices S.5 and Tables S.3, S.4 referenced in Section 5, and the R code referenced in Section 4 and 5 are available with this paper at the Biometrics website on Oxford Academic.

\section*{Data Availability}
The data that support the findings in this paper cannot be shared publicly due to constraints imposed by Data Use Agreements with Kaiser Permanente.


%




\label{lastpage}

\bibliographystyle{biom} 
\bibliography{bibilo.bib}
\end{document}